\documentclass{moriond}

\usepackage{color}
\definecolor{LinkCol}{RGB}{0,30,120}
\hypersetup{linktocpage,
    colorlinks,
    linkcolor={LinkCol},
    citecolor={LinkCol},
    urlcolor={LinkCol}
}

\usepackage{amsmath}
\usepackage{amssymb}
\usepackage{bbm}
\usepackage{slashed}
\usepackage[superscript]{cite}

\def\be{\begin{equation}}
\def\ee{\end{equation}}
\def\bea{\begin{eqnarray}}
\def\eea{\end{eqnarray}}

\newcommand{\cf}{\textit{cf.}}
\newcommand{\Photo}{}

\begin{document}
\vspace*{4cm}
\title{Flavored Dark Matter: Freeze-out Scenarios and LHC Signatures}

\author{Jan Heisig}

\address{Institute for Theoretical Particle Physics and Cosmology, RWTH Aachen University, \\ 
D-52056 Aachen, Germany}

\maketitle
\abstracts{
We present the phenomenology of a simplified dark matter model within the dark minimal flavor violation framework, extending the standard model with a Majorana fermion flavor triplet and colored scalar mediator. We explore the allowed parameter space considering constraints from flavor observables, direct and indirect dark matter detection, and the relic density measurement. Our analysis reveals previously unexplored scenarios such as conversion-driven freeze-out within the model. We study limits from the LHC and discuss new signatures of the model, highlighting its richer phenomenology compared to single flavor  dark matter.
}

\section{Introduction}

Understanding the underlying principles of matter is among the primary motivations for research in modern physics. Both the flavor structure of visible matter and the nature of Dark Matter (DM) have sparked extensive theoretical and experimental investigations. In this proceedings article, we summarize previous work\cite{Acaroglu:2023phy} on flavored DM within the Dark Minimal Flavor Violation (DMFV) hypothesis.\cite{Agrawal:2014aoa,Chen:2015jkt} This concept provides a simple yet powerful framework to study possible links between flavor physics and the physics of DM\@.

Similar to supersymmetry-inspired simplified DM models, flavored DM models extend the Standard Model (SM) by a gauge singlet DM and a charged or color-charged $t$-channel mediator, akin to the superpartners of leptons or quarks. However, unlike these simplified models, flavored DM models contain a DM flavor triplet. Accordingly, the Yukawa-type interaction between DM, the mediator, and the SM leptons or quarks involves a $3\times3$ coupling matrix in flavor space. Similar to the concept of minimal flavor violation, DMFV assumes that this coupling matrix is the only new source of flavor and CP violation beyond the SM Yukawa interactions. This assumption allows us to study the phenomenology of the particle spectrum at accessible energies while parametrizing the effects of possible UV-completions in an effective way.

Flavored DM has been studied in both quark-philic and lepto-philic scenarios, each with interesting implications. For instance, lepto-philic models have been examined in the context of the anomalous magnetic moment of the muon, providing a potential explanation for the long-standing (although currently scrutinized) discrepancy between the SM prediction and the experimental measurement.\cite{Acaroglu:2022boc} Generally, the extended flavor structure helps reconcile the tension between relic density measurements and constraints from DM searches within the paradigm of thermal freeze-out of a weakly interacting massive particle (WIMP).\cite{Blanke:2017tnb,Acaroglu:2021qae}

As we recently showed\cite{Acaroglu:2023phy} and report in this article, there are further ways to evade the strengthening constraints from WIMP searches. Specifically, flavored DM models provide various realizations of conversion-driven freeze-out.\cite{Garny:2017rxs,DAgnolo:2017dbv} In this scenario, the thermal decoupling of DM in the early Universe is not related to DM annihilation (as it is the case in canonical freeze-out) but to the conversion processes among the particles in the new physics sector. Consequently, the couplings required to explain the relic density are considerably weaker than those of WIMPs, typically of the order of $10^{-6}$, aligning with the null-results of direct and indirect detection experiments.\cite{Garny:2017rxs} Interestingly, this scenario predicts long-lived particles at colliders. Furthermore, it offers an intriguing link to baryogenesis when realized in lepton-flavored DM models.\cite{Heisig:2024mwr}

In this article, we focus on the quark-philic case, coupling the DM multiplet to the right-handed up-type quarks. We introduce the model in Sec.~\ref{sec:model}. In Sec.~\ref{sec:FOscenl}, we discuss the viable parameter space in the canonical and conversion-driven freeze-out regimes, considering astrophysical and flavor constraints. We then discuss current constraints from searches at the LHC and point out future opportunities for dedicated probes in Sec.~\ref{sec:LHC}, before concluding in Sec.~\ref{sec:conl}.

\section{Flavored Dark Matter Model}
\label{sec:model}

We consider a simplified model extending the SM by a Majorana fermion flavor triplet $\chi$ and a colored mediator $\phi$, which carries the same gauge quantum numbers as the up-type quarks. Both fields are assumed to be odd under a new $\mathbb{Z}_2$ symmetry, ensuring DM stabilization. The Lagrangian reads~\cite{Acaroglu:2021qae}:
\begin{eqnarray}
\label{eq:lagr}
    \mathcal{L}&=& \mathcal{L}_\text{SM} + \frac{1}{2}\left(i\bar\chi\slashed\partial\chi - M_\chi \bar\chi\chi\right)
    -\left(\lambda_{ij}\bar u_{Ri} \chi_j\phi + \text{h.c.}\right) \nonumber\\
    &&+(D_\mu \phi)^\dagger (D^\mu\phi) -m_\phi^2\phi^\dagger\phi  +\lambda_{H\phi}\phi^\dagger\phi  H^\dagger H +\lambda_{\phi\phi}\left(\phi^\dagger\phi \right)^2\,,
\end{eqnarray}
where $M_\chi$ is the Majorana mass matrix, $\lambda_{ij}$ is a complex $3\times3$ coupling matrix in quark and DM flavor space, $m_\phi$ is the mediator mass, and $\lambda_{H\phi}$ and $\lambda_{\phi\phi}$ are the Higgs-portal and quartic couplings of $\phi$, respectively.

Within the DMFV spurion expansion, $M_\chi$ can be written as:\cite{Acaroglu:2021qae}
\begin{equation}
\label{eq:masscorr}
M_\chi = m_\chi \left[\mathbbm{1} + \eta \,\text{Re}(\lambda^\dagger\lambda) + \mathcal{O}(\lambda^4) \right]\,,
\end{equation}
where $\eta$ parametrizes the contributions from higher-dimensional operators induced by the theory's UV completion. Note that a misalignment between the flavor and the mass eigenstates for $\chi$ can be absorbed in a field redefinition via an Autonne-Takagi factorization, rendering $M_\chi$ diagonal. Without loss of generality, we choose $m_{\chi_3}\le m_{\chi_2}\le m_{\chi_1}$. The flavor-violating coupling matrix obeys an $O(3)_\chi$ flavor symmetry, allowing for a parametrization of the form $\lambda = U D\, O\, d$, with a total of 15 physical parameters including mixing angles and complex phases, as well as three real non-negative coupling parameters $D_i$ entering the diagonal matrix $D$.\cite{Acaroglu:2021qae} 
The couplings $\lambda_{H\phi}$ and $\lambda_{\phi\phi}$ in Eq.~\eqref{eq:lagr} are not relevant for the considered phenomenology.\cite{Acaroglu:2023phy}

\section{Freeze-out Scenarios}
\label{sec:FOscenl}

Within the considered model, the measured DM relic density can be explained across a wide range of couplings utilizing different DM freeze-out scenarios in the early Universe. In the WIMP regime, canonical freeze-out, with or without coannihilation effects, provides cosmologically viable scenarios. For illustration, we distinguish between three sub-scenarios. In the Single Flavor Freeze-Out (SFF) scenario, all heavier new physics states ($\chi_1$, $\chi_2$, and $\phi$) are assumed to be at least 10\% heavier than $\chi_3$, rendering any coannihilation effects irrelevant. In contrast, in the Quasi-Degenerate Freeze-Out (QDF) scenario, we assume all $\chi_i$ to be quasi-mass degenerate, maximizing coannihilation among the DM multiplet states, while $m_\phi$ is still assumed to be heavier by 10\%. Finally, in Generic Canonical Freeze-Out (GCF), we drop any restriction on the mass hierarchy, only requiring $m_{\chi_3}$ to be the lightest among all new physics states.

\begin{figure}
	\centering
		\includegraphics[width=0.45\textwidth]{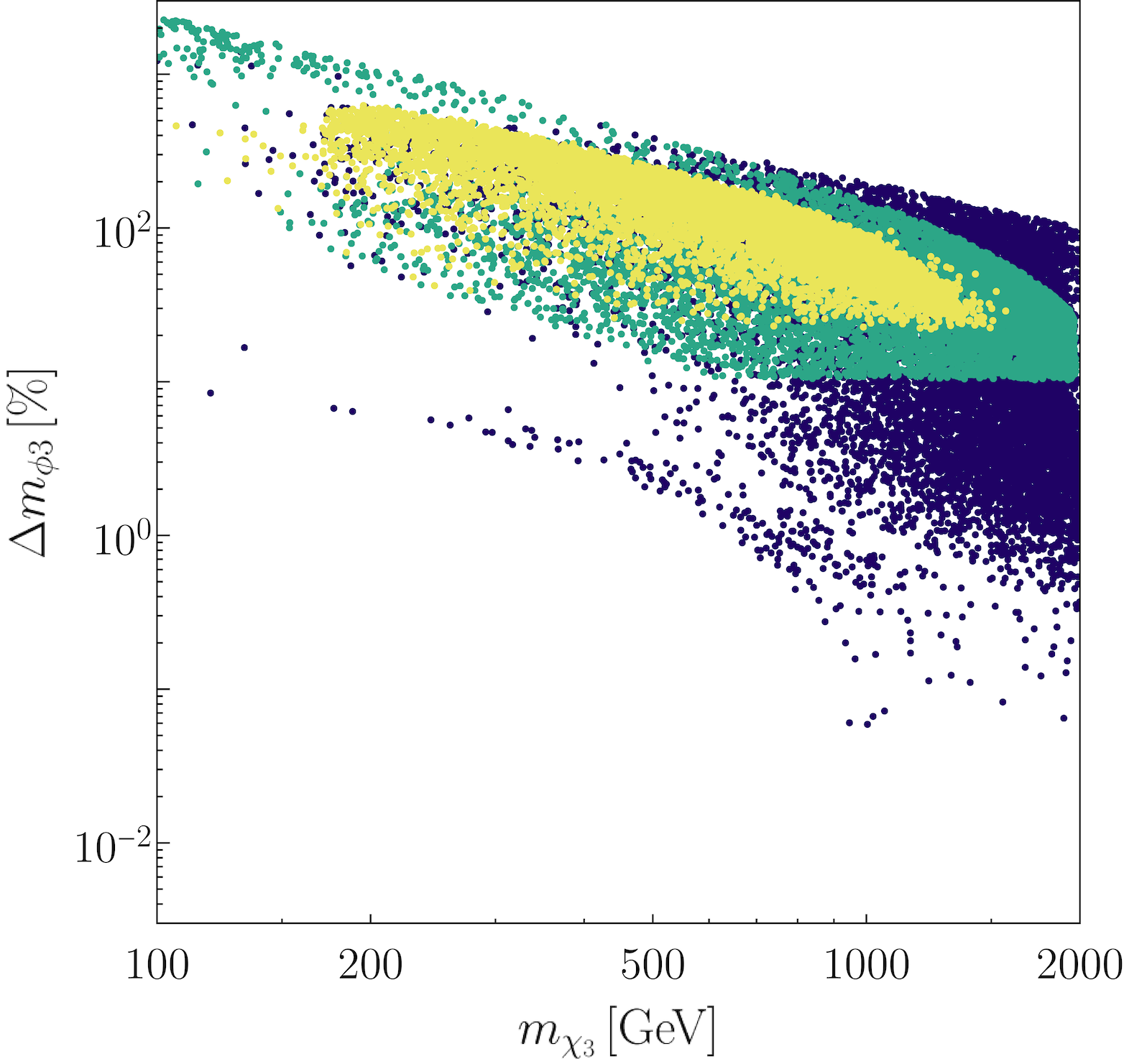}\hspace{2ex}
		\includegraphics[width=0.45\textwidth]{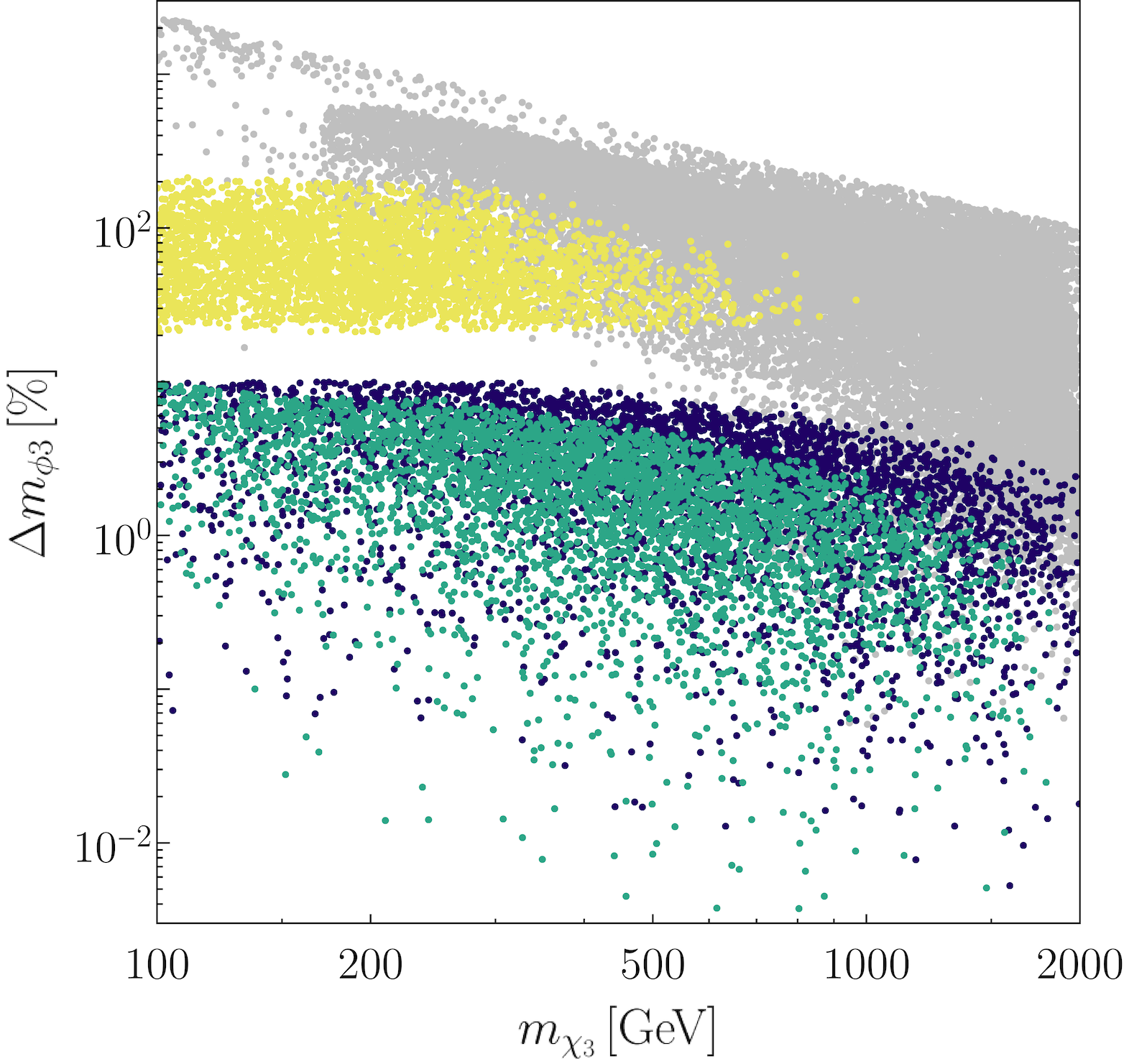}
	\caption{\emph{Left panel:} Allowed parameter space for the canonical freeze-out scenarios. The yellow, green, and blue points correspond to the SFF, QDF, and GCF scenarios, respectively.
\emph{Right panel:} Allowed parameter space in the conversion-driven freeze-out scenarios. Here, the yellow, green, and blue points correspond to C$_\chi1_u$, C$_\phi2_u$, and C$_\phi1_u$, respectively, while the light gray points show the canonical scenario for comparison. Taken from Ref.~{\color{LinkCol}1}.
	}
    \label{fig:FOsc}
\end{figure}

The left panel of Fig.~\ref{fig:FOsc} shows viable parameter points in the three sub-scenarios of canonical freeze-out. To obtain this result, we perform a Monte Carlo parameter scan imposing several constraints:\cite{Acaroglu:2023phy}
\vspace{-1ex}
\begin{itemize}\setlength\itemsep{-0.22em}
\item
Relic Density: We require $\Omega_{\mathrm{DM}} h^2 = 0.12$\cite{Planck:2018vyg} within a 10\% theoretical uncertainty.
\item
Flavor: We impose constraints from flavor observables, particularly $D^0$--$\bar{D}^0$ meson mixing.\cite{Acaroglu:2021qae}
\item
Direct Detection: We impose constraints on the spin-independent and spin-dependent scattering cross-sections from LZ.\cite{LZ:2022lsv}
\item
Indirect Detection: We apply limits from cosmic-ray antiproton flux measurements by AMS-02 on the DM annihilation cross-section.\cite{Cuoco:2017iax}
\end{itemize}
\vspace{-1ex}
The results indicate that the model can explain the DM relic density and evade DM detection and flavor constraints in a wide range of masses with and without significant coannihilation effects. This stands in contrast to the respective single flavor model in which the WIMP parameter space is almost fully excluded by the above constraints, \cf~the $t$-channel mediator model with the same gauge quantum numbers and spin assignments (`S3M$u_R$').\cite{Arina:2023msd}

If we allow for very weak couplings, further viable parameter space opens up. In the lower left corner of the left panel of Fig.~\ref{fig:FOsc}, DM turns out to be underabundant for WIMP-like couplings. This is due to the large mediator pair annihilation rate, which supplies too efficient a depletion of dark sector particles when in chemical equilibrium with each other. However, for very weak couplings $\lambda$, typically in the ballpark of $10^{-6}$, the chemical equilibrium can break down as conversion processes among dark sector particles become semi-efficient. In this case, conversions can initiate the freeze-out.\cite{Garny:2017rxs,DAgnolo:2017dbv} This conversion-driven freeze-out scenario provides an alternative cosmologically viable solution within the model, specifically in the region of small mass splitting. 

In the following, we consider three sub-scenarios. In C$_\phi1_u$ and C$_\phi2_u$, respectively, one and two DM multiplet states have very weak couplings while the mediator $\phi$ is close in mass, providing the dominant contribution to (co)annihilation. In C$_\chi1_u$, one DM multiplet state is very weakly coupled while now the heavier $\chi_i$ (instead of $\phi$) are responsible for (co)annihilation. In all cases, DM predominantly couples to the $u$-quark. The allowed points in the three sub-scenarios are shown in the right panel of Fig.~\ref{fig:FOsc}. Note that due to the small DM coupling, indirect and direct detection constraints do not challenge the parameter space. However, flavor constraints significantly affect the results, particularly in the C$_\chi1_u$ scenario which requires sizeable couplings for the heavier $\chi_i$ that drive the dark sector depletion through their pair-annihilation.

\section{Probes at the LHC}
\label{sec:LHC}

\subsection{Current LHC Constraints}

\begin{figure}
	\centering
		\includegraphics[width=0.45\textwidth]{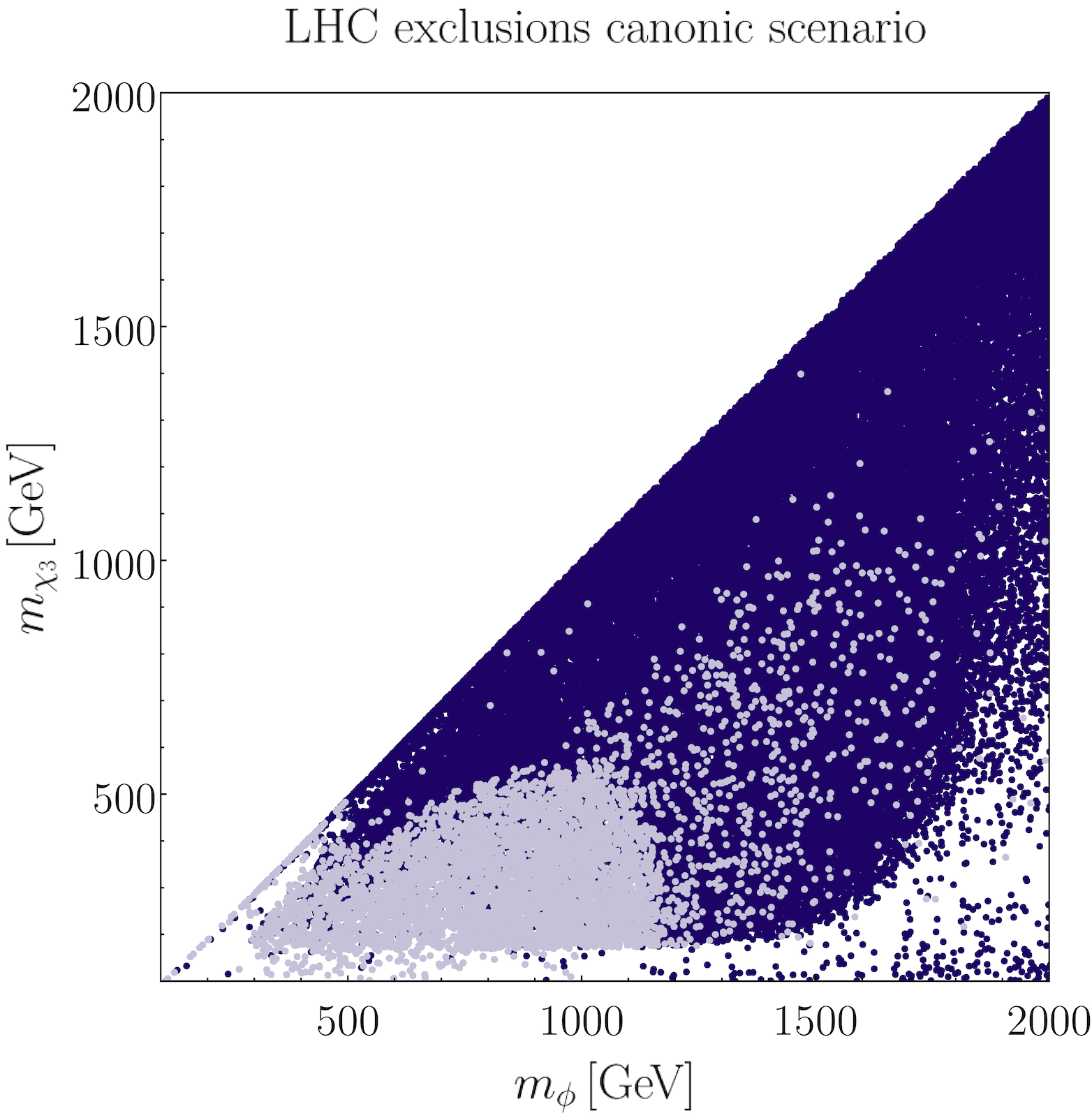}\hspace{2ex}
		\includegraphics[width=0.45\textwidth]{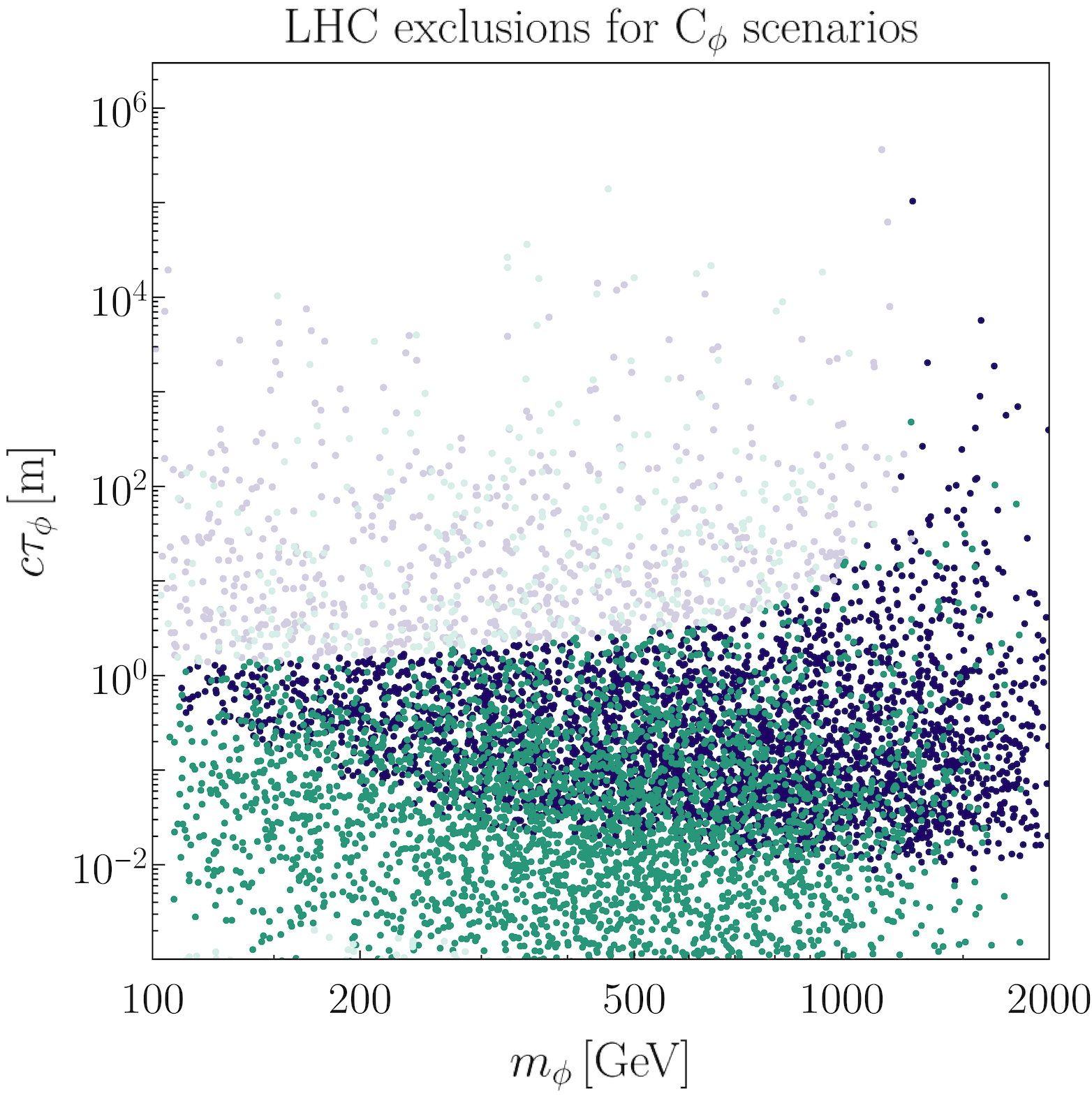}
        \caption{95\% CL exclusion limits from LHC searches. Points displayed in faint colors are excluded. \emph{Left panel:} Constraints from $\slashed{E}_T$ searches on the canonical freeze-out scenario. \emph{Right panel:} Constraints from long-lived particle searches on the conversion-driven freeze-out scenarios C$_\phi1_u$ (blue) and C$_\phi2_u$ (green). Taken from Ref.~{\color{LinkCol}1}.
        }
          \label{fig:LHC1}
\end{figure}

Similar to other $t$-channel mediator models, the considered flavored Majorana DM model can be probed by searches for new physics at the LHC\@. Due to its strong interactions, mediator pair production followed by its (cascade) decay into DM provides an important search channel. We use the program package \textsc{SModelS}\cite{Alguero:2021dig,MahdiAlt:2023bdn} for a convenient reinterpretation of a large number of LHC searches, mostly targeted at supersymmetric scenarios predicting similar signatures. The left panel of Fig.~\ref{fig:LHC1} shows the corresponding constraints considering the canonical freeze-out scenario (GCF including SFF and QDF). The light shaded points are excluded by searches for jets + $\slashed{E}_T$ and tops + $\slashed{E}_T$.\cite{ATLAS:2017mjy,ATLAS:2017drc,ATLAS:2020dsf,ATLAS:2020syg,CMS:2017abv,CMS:2017okm,CMS:2019zmd,CMS:2021eha}

For most points, the signature is similar to the simplified models the searches were originally interpreted in. Thus, most excluded (allowed) points lie within (beyond) the respective exclusion limits reported in the LHC searches. However, there are excluded points with significantly larger masses as well as points with very small masses that escape the exclusion limits. The former effect originates from a significant contribution to the mediator pair production from diagrams with DM multiplets in the $t$-channel, which can be sizable for large DM couplings. This particularly concerns the same-sign mediator production present for the Majorana multiplet under consideration, as it can be further enhanced by the contribution $uu \to \phi \phi$ levering the large valence $u$-quark content of the proton. Points with very small masses can escape the exclusion limits mostly for two reasons. First, the (complex) pattern of cascade decays does not match any results available in \textsc{SModelS}. This is mainly a technical limitation and a proper recasting of the analyses is expected to reveal the exclusion of these points. Secondly, some of the decays in the decay chain are non-prompt, challenging the application of the considered $\slashed{E}_T$. This case may point to actual gaps in the coverage of LHC searches.

In the conversion-driven freeze-out region, the very weak coupling of DM predicts long-lived particles at the LHC\@. In the right panel of Fig.~\ref{fig:LHC1}, we show the current constraints from long-lived particles on the scenarios C$_\phi1_u$ (blue) and C$_\phi2_u$ (green), again utilizing \textsc{SModelS}. In these scenarios, the mediator becomes metastable, leading to anomalous tracks, disappearing tracks, or displaced vertices. However, only the former signature provides constraints\cite{ATLAS:2019gqq,CMS-PAS-EXO-16-036} that are directly applicable to our scenario, constraining the points with decay lengths larger than around a meter. While a more careful recasting provides some constraining power of the latter search strategies towards smaller decay lengths, the relatively small mass splittings leading to soft displaced objects provides experimental challenges yet to be addressed.\cite{Heisig:2024xbh}

\subsection{Majorana-specific Signatures}

So far, we have not explicitly made use of the Majorana nature of the DM multiplet. However, the above-mentioned enhanced same-sign mediator production channel $pp \to \phi \phi$ provides various opportunities. First, this channel can lead to the prominent signature of same-sign tops and missing energy, potentially improving our constraints. However, recasting the recent CMS search for same-sign top signatures,\cite{CMS:2020cpy} we find that the sensitivity of jets + $\slashed{E}_T$ and tops + $\slashed{E}_T$ searches considered earlier\cite{Acaroglu:2021qae} is significantly stronger than those from the same-sign $ttjj + \slashed{E}_T$. This is partly due to the additional jets required in the considered CMS search and partly due to the requirement of two semi-leptonic top decays, reducing the overall signal.

\begin{figure}[t!]
	\centering
	\includegraphics[width=0.45\textwidth]{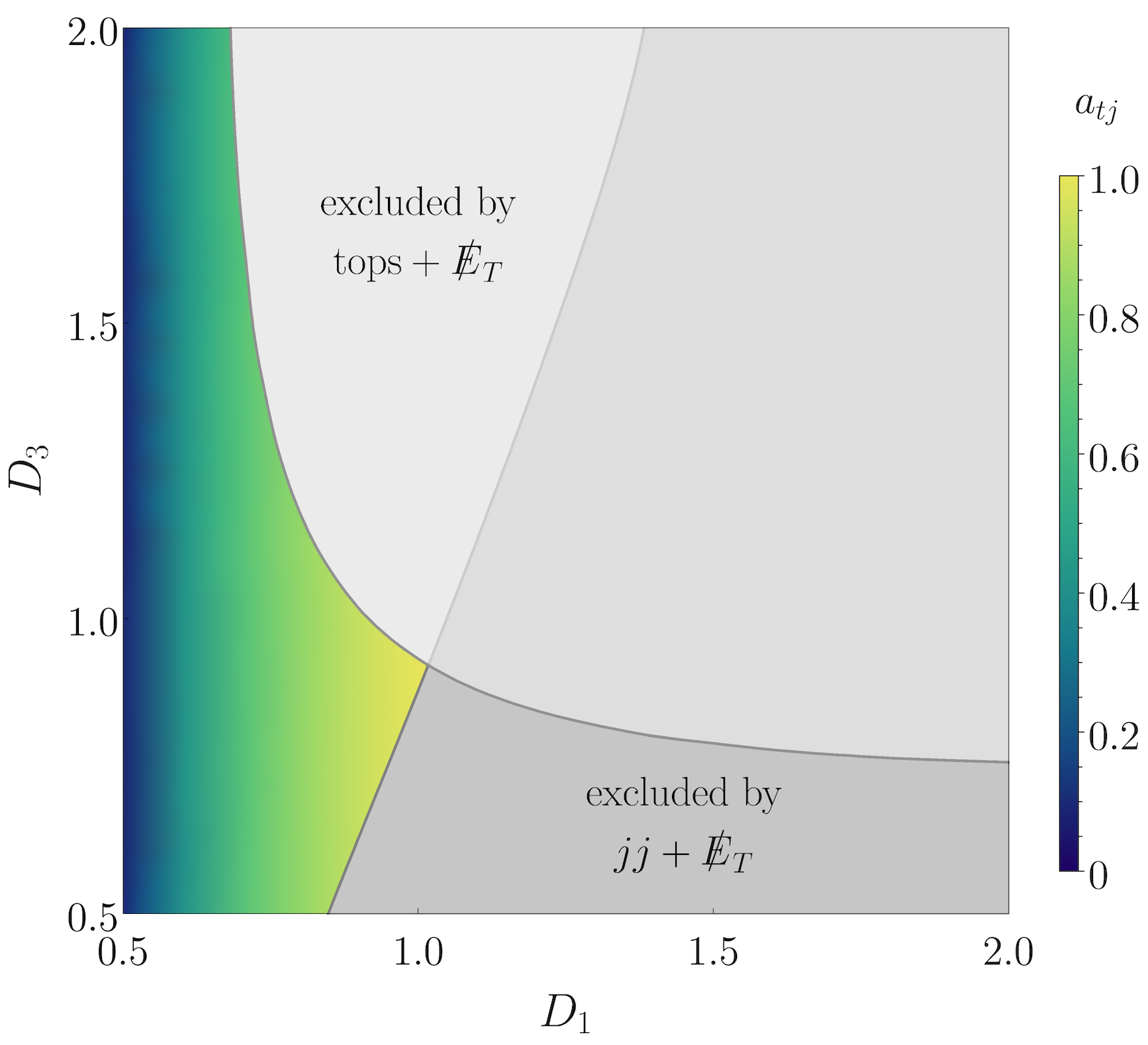}\hspace{2ex}
	\includegraphics[width=0.4036\textwidth]{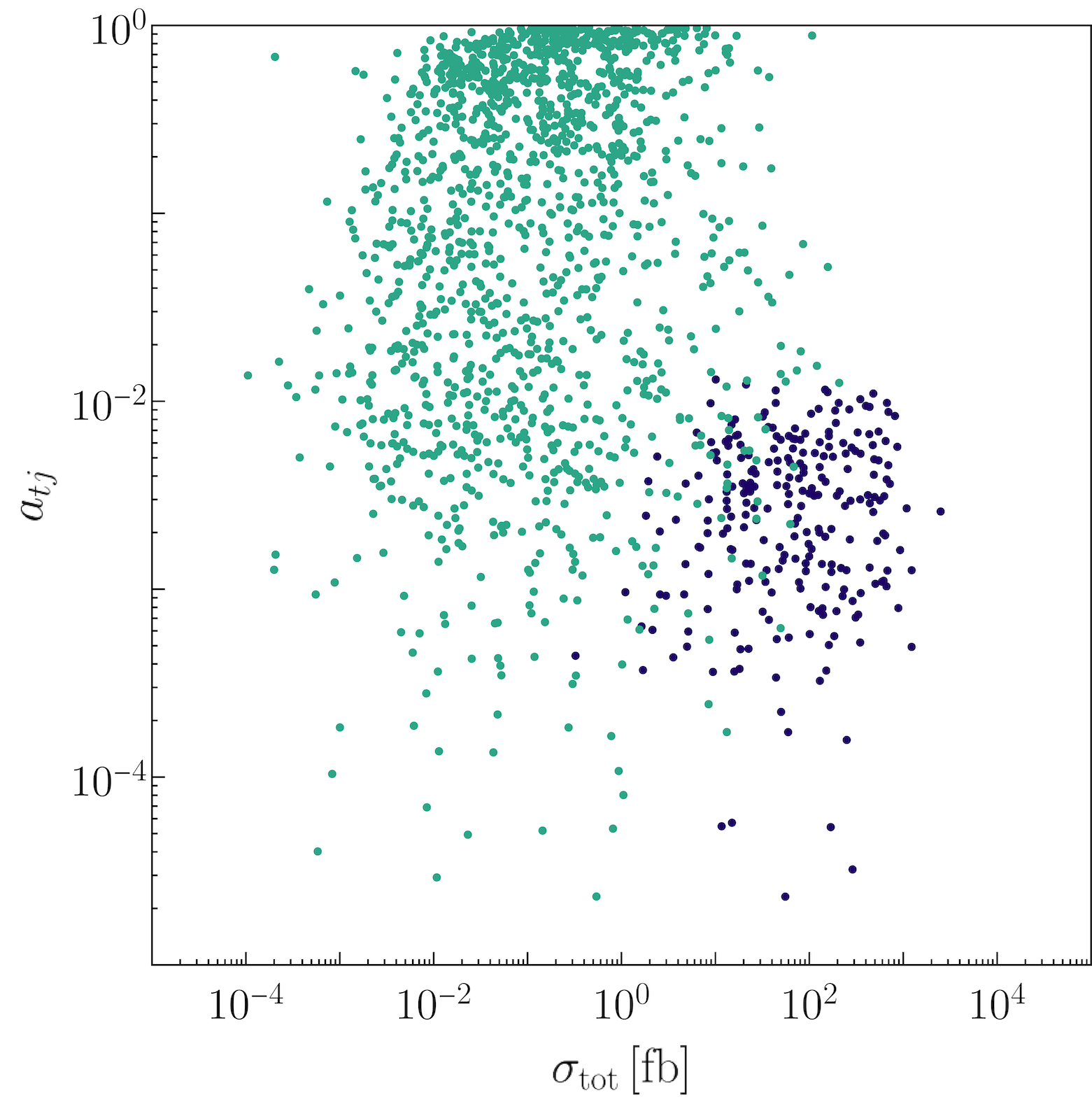}
 \caption{\emph{Left panel:} Predicted charge asymmetry, $a_{tj}$, (color code) at the 14\,TeV LHC and constraints from tops+$\slashed E_T$ (light gray) and jets+$\slashed E_T$ (dark gray)  for $D_2=0$, $m_\phi = 1200\,\mathrm{GeV}$, $m_\chi = 400\,\mathrm{GeV}$. \emph{Right panel:}~Predicted $a_{tj}$ and $\sigma_\text{tot}$ of the viable parameter space in the GCF (green) and C$_\chi 1_u$ scenario (blue). Taken from Ref.~{\color{LinkCol}1}.
 }
 \label{fig:LHC2}
\end{figure}

A second way to explore the Majorana nature of DM is to consider mediator pair production with subsequent decays into $j + \slashed{E}_T$ and $t + \slashed{E}_T$, respectively. This process induces the flavor-violating final state $tj + \slashed{E}_T$. Interestingly, due to the enhanced same-sign mediator production channel, $\sigma(tj + \slashed{E}_T) > \sigma(\bar{t}j + \slashed{E}T)$, unlike in the case of Dirac DM\@. Accordingly, we can consider the charge asymmetry
\begin{equation}
    a_{tj} = \frac{\sigma(tj+\slashed E_T)-\sigma(\bar tj+ \slashed E_T)}{\sigma(tj+\slashed E_T)+\sigma(\bar tj+\slashed E_T)}\;\;
    \begin{cases} 
   \;\; > 0 & \mathrm{Majorana} \\ 
   \;\; \simeq 0 & \mathrm{Dirac}
    \end{cases}\,.
\end{equation}
The left panel of Fig.~\ref{fig:LHC2} shows $a_{tj}$ as a function of $D_1$ and $D_3$, which parametrize the couplings of the DM multiplet states to the first and third generation of quarks, respectively, only the former of which affects the cross section for $uu \to \phi \phi$. For comparison, we show current limits from jets + $\slashed{E}_T$ and tops + $\slashed{E}_T$.\cite{Acaroglu:2023phy}

The right panel of Fig.~\ref{fig:LHC2} displays $a_{tj}$ and the total cross section $\sigma_\text{tot}=\sigma(tj+\slashed E_T)+\sigma(\bar tj+\slashed E_T)$ for a subset of points within the canonical and conversion-driven freeze-out scenarios. The former (green points) provides cross sections up to $\sigma_\text{tot} \simeq 100,\text{fb}$ for $\mathcal{O}(1)$ values of $a_{tj}$, rendering the charge asymmetry a promising observable to probe the nature of DM\@. For the conversion-driven freeze-out scenario C$_\chi 1_u$, in contrast, $a_{tj}\lesssim10^{-2}$. Hence, a measurement different from zero would impose strong constraints on this scenario.

\section{Conclusions}
\label{sec:conl}

We studied a DM model within the Dark Minimal Flavor Violation framework, featuring a Majorana fermion flavor triplet and a colored scalar mediator. We identified cosmologically viable freeze-out scenarios, exploring coannihilation effects as well as conversion-driven freeze-out with a very weak DM coupling. We confronted the scenarios with flavor constraints, particularly from $D^0-\bar{D}^0$ mixing, and DM searches, revealing a wide range of viable DM masses, in contrast to the corresponding single flavor \(t\)-channel mediator model. 

Using \textsc{SModelS}, we assessed LHC constraints, finding exclusions from jets + \(\slashed{E}_T\) and tops + \(\slashed{E}_T\) searches, as well as long-lived particles in the canonical as well as conversion-driven freeze-out scenario, respectively. The latter case revealed significant gaps in the coverage of LHC searches for soft displaced objects. Additionally, we highlighted two model-specific signatures to probe the Majorana nature of the DM multiplet: same-sign di-top + \(\slashed{E}_T\) and single-top charge asymmetry. Both signature make use of the enhanced $t$-channel contribution $uu\to \phi \phi$ to the mediator production. Our analysis showed that a non-zero single-top charge asymmetry could indicate the Majorana nature of DM, rendering it a promising observable for future LHC studies.

\section*{Acknowledgments}
I thank Harun Acaro\u{g}lu, Monika Blanke, Michael Kr\"amer, and Lena Rathmann for fruitful collaboration. I acknowledge support from the Alexander von Humboldt Foundation via the~Feodor Lynen Research Fellowship for Experienced Researchers and Feodor Lynen Return Fellowship.

\section*{References}

\bibliographystyle{morUnsrt}

\end{document}